\documentclass[twocolumn,prb]{revtex4}
\usepackage[dvips]{graphicx}
\usepackage{amscd}
\usepackage{xspace}
\usepackage{amstext}
\usepackage{amssymb,amsfonts,amsthm}
\usepackage{amscd}
\usepackage{hyperref}

\newcommand{\dd}{{\rm d}}

\newtheorem{theorem}{Theorem}[section]

\begin{document}


\title{Differential aging from acceleration, an explicit formula}

\author{E. Minguzzi}
 \affiliation{Departamento de Matem\'aticas,
Universidad de Salamanca, \\ Plaza de la Merced 1-4, E-37008
Salamanca, Spain \\ and INFN, Piazza dei Caprettari 70, I-00186
Roma, Italy}
\email{minguzzi@usal.es }

\begin{abstract}
We consider a clock ``paradox" framework where an observer leaves
an inertial frame, is accelerated and after an arbitrary trip
comes back. We discuss a simple equation that gives, in the 1+1
dimensional case, an explicit relation between the time elapsed on
the inertial frame and the acceleration measured by the
accelerating observer during the trip.

A non-closed trip with respect to an inertial frame appears closed
with respect to another suitable inertial frame. Using this
observation we define the differential aging as a function of
proper time and show that it is non-decreasing.  The
reconstruction problem of special relativity is also discussed
showing that its, at least numerical, solution would allow the
construction of an {\em inertial clock}.
\end{abstract}

 \maketitle

\section{Introduction}
The differential aging implied by special relativity is surely one
of the most  astonishing results of modern physics (for an
historical introduction see\cite{pesic03}, for a bibliography with
old papers see\cite{uno}).
 It has been largely debated,
 and in particular the relationship between the role of acceleration
and the difference in proper times of inertial and accelerated
observers has been discussed\cite{due}. The old question as to
whether acceleration could be considered responsible for
differential aging receives a simple answer by noticing that
proper and inertial time are related in the time dilation effect;
since relative velocity enters there so does acceleration changing
the velocity. The acceleration, however,  is not the ultimate
source of differential aging as the twin paradox in non-trivial
spacetime topologies can be reformulated without any need of
accelerated observers \cite{tre}.

Here we give a simple equation that relates acceleration and
differential aging in  the case the accelerated observer undergoes
an unidirectional, but otherwise arbitrary, motion. We shall prove
that relation in the next section. Here we want to discuss and
apply it to some cases previously investigated with more
elementary methods.

Choose units such that $c=1$. Let $K$ be the inertial frame and
choose coordinates in  a way such that two of them can be
suppressed. Let $O$ be an accelerated observer with timelike
worldline $x^{\mu}: [0, \bar\tau] \to M$  and
$x^{1}(0)=x^{1}(\bar\tau)=0$, where $\tau$ is the proper time
parametrization and $x^{\mu}$, $\mu=0,1$ are the coordinates of
the inertial frame. Let, moreover, $a(\tau)$ be the acceleration
of $O$ with respect to the local inertial frame at $x(\tau)$.  To
be more precise, the quantity $-a$ is the apparent acceleration
measured by $O$ and so it has a positive or negative sign
depending on the direction. Let $T=x^{0}(\bar\tau)-x^{0}(0)$ be
the (positive) inertial time interval between the departure and
arrival of $O$, we have \footnote{These theorems hold also in a
non-topologically  trivial Minkowski spacetime but there $O$
should move along a worldline which is homotopic to $K$'s origin
worldline; this follows from their validity in the covering
Minkowski spacetime.}

\begin{theorem}
The time dilation $T$ is related to the acceleration $a(\tau)$ by
(time dilation-acceleration equation)
\begin{equation}\label{itf}
T^{2}=\left[ \int^{\bar\tau}_{0} e^{\int^{\tau}_{0} a(\tau')\dd
\tau'} \, \dd \tau\right] \,\left[\int^{\bar\tau}_{0}
e^{-\int^{\tau}_{0} a(\tau')\dd \tau'} \, \dd \tau \right].
\end{equation}
\end{theorem}
We have also
\begin{theorem}
The accelerated observer departs from $K$ with zero velocity if
and only if $\int^{\bar\tau}_{0} e^{\int^{\tau}_{0} a(\tau')\dd
\tau'} \, \dd \tau=\int^{\bar\tau}_{0} e^{-\int^{\tau}_{0}
a(\tau')\dd \tau'} \, \dd \tau$ and in this case
\begin{equation} \label{itf2}
T=\int^{\bar\tau}_{0} e^{\pm\int^{\tau}_{0} a(\tau')\dd \tau'} \,
\dd \tau,
\end{equation}
if moreover the final velocity of $O$ with respect to $K$ vanishes
then $\int^{\bar\tau}_{0} a(\tau)\dd \tau=0$.
\end{theorem}
Some comments are in order. In no place we need to specify the
initial or final velocity of $O$ with respect to $K$. Using the
Cauchy-Schwarz inequality $(\int fg \dd \tau)^{2}\le (\int f^{2}
\dd \tau ) (\int g^{2} \dd \tau)$, with
$f=g^{-1}=\exp(\int^{\tau}_{0} a(\tau')\dd \tau'/2)$ we find the
expected relation $T \ge \bar\tau$ where the equality holds only
if $f=kg$, with $k \in \mathbb{R}$, that is if and only if
$a(\tau)=0$. Thus $T>\bar\tau$ or the worldline of $O$ coincides
with that of the origin of $K$. This proves the differential aging
effect. In section \ref{boo} we shall give another proof that does
not use the Cauchy-Schwarz inequality.

Often \cite{moller62} the differential aging effect is proved in
curved (and hence even in flat) spacetimes by noticing that the
connecting geodesic, that is the trajectory of equation
$x^1(\tau)=0$ in our case, locally maximizes the proper time
functional $I[\gamma]=\int_\gamma {\dd \tau}$.  Theorem \ref{itf}
implies the global maximization property in 1+1 Minkowski
spacetime and has the advantage of  giving an explicit formula for
the inertial round-trip dilation.

\subsection{The simplest example} The simplest example is that of
uniform motion in two intervals $[0,\bar\tau/2]$ and
$[\bar\tau/2,\bar\tau]$. In the first interval $O$ moves with
respect to $K$ at velocity $v=\dd x^{1}/\dd x^{0}$, in the second
interval at velocity $-v$. Although this is a quite elementary
example it is interesting to look at the time
dilation-acceleration equation and see how it predicts the same
result. The first problem is that  Eq. (\ref{itf}) holds for
integrable acceleration functions. In this example, instead, the
acceleration has a singularity at $\bar\tau/2$ (the initial and
final singularities are not present if the motion of $O$ is not
forced to coincide with that of $K$'s origin for $\tau$ outside
the interval). The reader can easily check (or see next section),
that if $\theta(\tau)=\tanh^{-1}v(\tau)$ is the rapidity then
$\frac{\dd \theta }{\dd \tau}=a$ (this follows from the additivity
of the rapidity under boosts and the fact that a small increment
in rapidity coincides with a small increment in velocity with
respect to the local inertial frame) and so
\[
\Delta \theta=\int a \dd \tau .
\]
If the acceleration causes, in an arbitrary small interval
centered at $\tilde\tau$, a variation $\Delta \theta$ in rapidity
then we must write $a=\Delta \theta \delta(\tau-\tilde\tau)$ and
generalize the time dilation-acceleration equation with this
interpretation. In presence of such singularities, however, it is
no longer true that $T$ does not depend on the initial and final
velocities of $K$. Indeed, we need to use this information to find
the coefficient $\Delta \theta$. In the case at hand we have
\[
\Delta \theta=\tanh^{-1}(-v)-\tanh^{-1}v=-2\tanh^{-1}v .
\]
Inserting $a=-2\tanh^{-1}v\, \delta(\tau-\tilde\tau)$ in Eq.
(\ref{itf}) we find, after some work with hyperbolic functions
that, $T=\bar\tau/\sqrt{1-v^{2}}$ as expected. The reader should
not be surprised by the fact that this simple case needs so much
work, as this is a rather  pathological case. No real observer
would survive an infinite acceleration. The advantage of the time
dilation-acceleration equation turns out in more realistic cases.

\subsection{The constant acceleration case} This case has also been
treated extensively in the literature \cite{good82,desloge87}. The
hypothesis is that in the interval $[0,\bar\tau]$ we have $a=-g$
with $g \in \mathbb{R}$. Equation (\ref{itf}) gives immediately
\[
T^{2}=\left[ \int^{\bar\tau}_{0} e^{-g \tau} \, \dd \tau\right]
\,\left[\int^{\bar\tau}_{0} e^{g \tau} \, \dd \tau
\right]=\frac{2}{g^{2}}(\cosh g\bar\tau-1)
\]
or $T=\frac{2}{ g}\sinh\frac{g\bar\tau}{2}$.

\subsection{A more complicated example} This example was considered
by Taylor and Wheeler \cite{taylor66}. It has the advantage that
the acceleration has no Dirac's deltas and $O$ departs from and
arrives at $K$ with zero velocity. The interval is divided into
four equal parts of proper time duration $\bar\tau/4$. The
acceleration in these intervals is successively  $g$, $-g$, $-g$
and $g$.

One can easily convince him/herself  that since the acceleration
in the second half interval is opposite to the one in the first
half interval the observer indeed returns to $K$'s worldline.
Moreover, we know that $O$ starts with zero velocity so we can
apply equation (\ref{itf2}). First we have
\begin{displaymath}
\int_{0}^{\tau} a(\tau') \dd \tau' =\left\{
\begin{array}{ll}
g\tau, & \ \tau \in [0, \frac{1}{4}\bar\tau], \\
-g\tau+g\bar\tau/2, & \ \tau \in [\frac{1}{4}\bar\tau,
\frac{3}{4}\bar\tau], \\
g\tau-g\bar\tau, & \ \tau \in [\frac{3}{4}\bar\tau, \bar\tau].
\end{array} \right.
\end{displaymath}
Integrating simple exponentials we arrive at $T=\frac{4}{
g}\sinh\frac{g\bar\tau}{4}$.

\begin{figure}[!t]
\centering
\includegraphics[width=6cm]{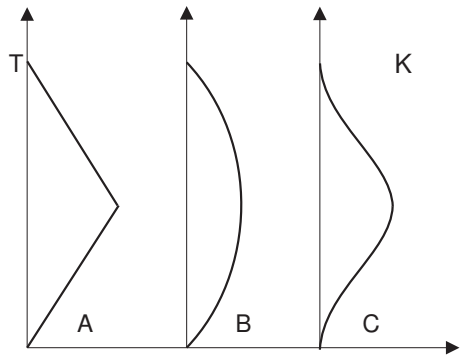}
\caption{The textbook round-trip examples.} \label{fig1}
\end{figure}

\section{The reconstruction problem in special relativity}
In this section we consider the problem of reconstructing the
motion in the inertial frame starting from the knowledge of the
acceleration. Similar mechanical problems have been studied
in\cite{rec}.
 It can be stated in  full Minkowski spacetime as follows.

{\em Consider a timelike worldline $x^{\mu}(\tau)$ on Minkowski
spacetime and a Fermi-transported triad $e_{i}$. Let
$a^{i}(\tau)=-(a(\tau)\cdot e_{i})$ be the components of the
acceleration vector with respect to the triad, $a=a^{i}e_{i}$.
Determine, starting from the data $a^{i}(\tau)$, the original
curve up to an affine transformation of Minkowski spacetime.}

Here the Fermi-transported triad represents gyroscopes. The
components of the acceleration with respect to this triad are
therefore measurable by $O$ using three orthogonal gyroscopes and
an accelerometer. The solution to this problem may be relevant for
future space travellers. Indeed, although the twin `paradox' has
been studied mainly assuming the possibility of some communication
by light signals, it is more likely that when distances grow
communication becomes impossible. Suppose the space traveller does
not want to be lost but still wants the freedom to choose time by
time its trajectory, then he/her should find some way to know its
inertial coordinates. The only method, if no references in space
are given, is to solve the reconstruction problem. Keeping track
of the acceleration during the journey the observer would be able
to reconstruct its inertial coordinates without looking outside
the laboratory. In particular he would be able to construct
(merging an accelerometer, three gyroscopes, and an ordinary
clock)  an {\em inertial clock} i.e. a clock that displays
$x^{0}(\tau)$.

The solution to the reconstruction problem  gives also to $O$ the
advantage of knowing its own position even before $K$ knows it.
Indeed, $O$ can know $x^{\mu}(\tau)$ immediately while $K$ has to
wait for light signals from $O$. In case of perturbations in the
trajectory, $O$ can immediately apply some corrections while, for
great distances, a decision from $K$ would take too much time.

In 1+1 dimensions the reconstruction problem can be solved easily.
For higher dimensions it becomes much more complicated and
numerical methods should be used. Let us give the solution to the
1+1 case. We use the timelike convention $\eta_{00}=1$.

If $v^{\mu}=\dd x^{\mu}/{\dd x^{0}}$, $v=\dd x^{1}/\dd x^{0}$ and
$u^{\mu}=\dd x^{\mu} /\dd \tau$, then we have
\[
a^{\mu}=\frac{\dd u^{\mu}}{\dd \tau}=\frac{\dd}{\dd
\tau}\frac{v^{\mu}}{\sqrt{1-v^{2}}}.
\]
Let $(0,a)$ be the components of the acceleration in the local
inertial frame (the first component vanishes because $u\cdot
a=0$). Since the square of the acceleration is a Lorentz invariant
we have $-a^{2}=a^{\mu}a_{\mu}$ or
\begin{eqnarray*}
-a^{2}&=&(\frac{\dd}{\dd
\tau}\frac{1}{\sqrt{1-v^{2}}})^{2}-(\frac{\dd}{\dd
\tau}\frac{v}{\sqrt{1-v^{2}}})^{2}\\
&=&-\frac{1}{(1-v^{2})^{2}}(\frac{\dd v}{\dd \tau})^{2} ,
\end{eqnarray*}
but $a$ has the same sign  as $\dd v/\dd \tau$ and hence
$a=\frac{\dd \theta}{\dd \tau}$, where $\theta=\tanh^{-1} v$ is
the rapidity, or
\begin{equation} \label{v}
v(\tau)=\tanh[\int^{\tau}_{0}  a(\tau') \dd \tau'+\tanh^{-1}v(0)].
\end{equation}
From $\dd x^{0}=\dd \tau/\sqrt{1-v^{2}}$ and $\dd x^{1}=\dd \tau\,
v /\sqrt{1-v^{2}}$ we have
\begin{eqnarray}
x^{0}(\tau)\!-\!x^{0}(0)\!\!\!&=&\!\!\!\!\!\int^{\tau}_{0}\!\!\!
\cosh[\int^{\tau'}_{0}\!\!\!\!\!\!
a(\tau'')\dd \tau''\!\!\!+\tanh^{-1}v(0)]\dd \tau', \\
x^{1}(\tau)\!-\!x^{1}(0)\!\!\!&=&\!\!\!\!\!\int^{\tau}_{0}\!\!\!
\sinh[\int^{\tau'}_{0} \!\!\!\!\!\!a(\tau'')\dd
\tau''\!\!\!+\tanh^{-1}v(0)]\dd \tau' , \label{pro}
\end{eqnarray}
Note that $v(0)$ is also easily measurable by $O$ since at
$\tau=0$, $K$ and $O$ are crossing each other. Without knowing
$v(0)$ the inertial coordinates are determined only up to a global
affine transformation. Indeed, we may say that the knowledge of
$v(0)$ specifies,  up to translations, the inertial coordinates
and frame with respect to which we  describe $O$'s motion.

Now, consider the  invariant under affine transformations
\begin{equation} \label{inv}
T^{2}(\tau)= [x^{0}(\tau)-x^{0}(0)]^{2}-[x^{1}(\tau)-x^{1}(0)]^{2}
.
\end{equation}
Since $x(\tau)$ is in the chronological future of $x(0)$ there is
a timelike geodesic passing through them. The inertial observer
$K(\tau)$ moving along that geodesic sees the motion of the
accelerated observer as a round trip. If $x_{K(\tau)}^{\mu}$ are
its coordinates $x_{K(\tau)}^{1}(\tau)=x_{K(\tau)}^{1}(0)=0$, thus
the previous invariant reads
\[
T(\tau)= x_{K(\tau)}^{0}(\tau)-x_{K(\tau)}^{0}(0) ,
\]
that is, $T(\tau)$ is the  travel duration with respect to an
inertial observer that sees the motion of the accelerated observer
as a round trip that ends at $\tau$. Using the relation
$a^{2}-b^{2}=(a-b)(a+b)$ we have from Eq. (\ref{inv})
\[
T^{2}(\tau)=\left[ \int^{\tau}_{0} e^{\int^{\tau'}_{0}
a(\tau'')\dd \tau''} \, \dd \tau'\right] \left[\int^{\tau}_{0}
e^{-\int^{\tau'}_{0} a(\tau'')\dd \tau''} \, \dd \tau'\right]. \]
Remarkably the dependence on $v(0)$ disappears. This follows from
the fact that contrary to $x^{0}(\tau)$ and $x^{1}(\tau)$, the
quantity $T(\tau)$ is a Lorentz invariant and as such should not
depend on the choice of frame (i.e. the  choice of $v(0)$).

In order to prove the second theorem note that if, with respect to
$K$, $O$ departs with zero velocity then from (\ref{pro}), after
imposing the round-trip condition $x^{1}(\bar\tau)=x^{1}(0)$, we
have
\[
\int^{\bar\tau}_{0} \sinh[\int^{\tau}_{0} a(\tau')\dd \tau'] \dd
\tau=0. \]
 that is, the two factors in the formula for $T^{2}$
coincide. Finally, if $O$ departs and returns with zero velocity
we have $\int_{0}^{\bar\tau} a \dd \tau=0$ as it follows from the
already derived relation $a=\dd \theta/\dd \tau$.

\subsection{Differential aging} \label{boo}

We give now a different proof that $T(\bar{\tau}) > \bar{\tau}$
unless $a(\tau)=0$ for all $\tau \in [0,\bar{\tau}]$ in which case
$T(\bar{\tau}) = \bar{\tau}$ and $O$ is  at rest in $K$.

The idea is to define the differential aging even for proper times
$\tau < \bar{\tau}$ as the differential aging between $K(\tau)$
and $O$. The differential aging at $\tau$ is therefore by
definition $\Delta(\tau) = T(\tau)-\tau$, that is the difference
between the proper time elapsed for an inertial observer that
reach $x(\tau)$ from $x(0)$ and that elapsed in the accelerating
frame. Roughly speaking if at proper time $\tau$ the accelerating
observer asks ``What is the differential aging now?" the answer
using this idea would be: it is the differential aging between you
and an imaginary twin who reached the same event where you are
now, but moving along a geodesic. This definition has the
advantage of avoiding conventions for distant simultaneity.

\begin{theorem}
The differential aging $\Delta(\tau)$ is a non-decreasing function
\begin{equation} \label{nd}
\frac{\dd \Delta}{\dd \tau} \ge 0 ,
\end{equation}
where the equality holds for all $\tau' \in [0,\tau]$ iff
$a(\tau')=0$ for all $\tau' \in [0,\tau]$.
\end{theorem}

\begin{proof}
Let $\Theta(\tau)=\int_{0}^{\tau} a(\tau) \dd \tau$. The
derivative of $T(\tau)$ is
\[
\frac{\dd T}{\dd \tau}=\cosh A(\tau) ,
\]
where
\[
A(\tau)= \Theta(\tau)+ \frac{1}{2} \ln \large\{
\frac{\int_{0}^{\tau} e^{-\Theta(\tau')} \dd
\tau'}{\int_{0}^{\tau} e^{\Theta(\tau')} \dd \tau'} \large\} .
\]
Since $\cosh A \ge 1$ this proves Eq. (\ref{nd}). Now, suppose
$\frac{\dd \Delta}{\dd \tau}(\tau)=0$ then $A(\tau)=0$ or
\begin{equation} \label{temp}
e^{-2\Theta}=\frac{\int_{0}^{\tau}e^{-\Theta(\tau')} \dd
\tau'}{\int_{0}^{\tau}e^{\Theta(\tau')} \dd \tau'}.
\end{equation}
Assume $\frac{\dd \Delta}{\dd \tau}(\tau')=0$ for all $\tau' \in
[0,\tau]$ then the previous equation holds for all $\tau'
 \in [0,\tau]$. Differentiating we obtain
\[
-2a(\tau)e^{-2\Theta}=\frac{e^{-\Theta}\int_{0}^{\tau}e^{\Theta(\tau')}\dd
\tau'-e^{\Theta}\int_{0}^{\tau}e^{-\Theta(\tau')}\dd
\tau'}{(\int_{0}^{\tau}e^{\Theta(\tau')}\dd \tau')^{2}}=0 ,
\]
that is $a(\tau)=0$ for all $\tau' \in [0,\tau]$.
\end{proof}

Since $\Delta(0)=0$ this theorem implies that $\Delta(\tau)>0$ for
$\tau>0$  unless $a(\tau')=0$ for all $\tau' \le \tau$. This
proves again the differential aging effect. However, the theorem
says something more. It proves that the definition of differential
aging we have given is particularly well behaved. It allows us to
say that, as proper time passes, the imaginary twin  is getting
older and older with respect to the accelerating observer.

\section{Conclusions}
We have discussed the reconstruction problem in special relativity
showing its relevance for the construction of {\em inertial
clocks} and in general for the positioning of the space traveller.
We have given a simple formula that relates the round-trip
inertial time dilation with the acceleration measured by the
non-inertial observer and have applied  it to some well know cases
to show how it works even in the presence of singularities. We
believe that it could be useful in order to explain clearly the
relationship between acceleration and differential aging
$T(\tau)-\tau$. Indeed, the differential aging effect is obtained
 easily by applying the Cauchy-Schwarz inequality.

Although there is a section on the twin paradox in almost every
textbook on special relativity, examples with singularities are
not always completely satisfactory, while more refined examples
require a lot of work. On the contrary the derivation of the time
dilation-acceleration formula is quite elementary needing only
some concepts from calculus. Its derivation as a classroom
exercise would probably convince students of the reality of the
differential aging effect.

\begin{acknowledgments}
I would like to acknowledge useful conversations with:  D. Amadei,
A. L\`opez Almorox, C. Rodrigo and C. Tejero Prieto. I am also
grateful to A. Macdonald for his comments and suggestions.
This work has been supported by INFN, grant $\textrm{n}^{\circ}$ 9503/02.\\
\end{acknowledgments}


\end{document}